\documentclass[10pt,usletter,conference]{IEEEtran}
\usepackage[top=2.5cm, bottom=2.0cm, left=1.45cm, right=1.45cm]{geometry}

\usepackage{graphicx}
\usepackage{subfigure} 
\usepackage{multirow}
\usepackage{makecell}
\usepackage[absolute,showboxes]{textpos}
\usepackage{hyperref}

\TPMargin{5pt}

\newcommand{\copyrightstatement}{
    \begin{textblock}{0.84}(0.08,0.95)    
         \noindent
         \footnotesize
         \copyright 2019 IEEE. Personal use of this material is permitted. Permission from IEEE must be obtained for all other uses, in any current or future media, including reprinting/republishing this material for advertising or promotional purposes, creating new collective works, for resale or redistribution to servers or lists, or reuse of any copyrighted component of this work in other works.
         DOI: \href{https://ieeexplore.ieee.org/abstract/document/8993477}{10.1109/IEDM19573.2019.8993477}
    \end{textblock}
}

\newcommand{\STAB}[1]{\begin{tabular}{@{}c@{}}#1\end{tabular}}

\begin{document}

\copyrightstatement

\title{\huge \textit{Ab initio} mobility of mono-layer MoS$_2$ and
  WS$_2$: comparison to experiments and impact on the device
  characteristics}

\author{\authorblockN{%
Y.~Lee, S.~Fiore, and M.~Luisier}
\authorblockA{Integrated Systems Laboratory, ETH Zurich, Zurich,
  Switzerland, email: youslee@iis.ee.ethz.ch}}

\maketitle

\begin{abstract}
We combine the linearized Boltzmann Transport Equation (LBTE) and
quantum transport by means of the Non-equilibrium Green's Functions
(NEGF) to simulate single-layer MoS$_2$ and WS$_2$ ultra-scaled
transistors with carrier mobilities extracted from experiments. 
Electron-phonon, charged impurity, and surface optical phonon
scattering are taken into account with all necessary parameters
derived from \textit{ab initio} calculations or measurements, except
for the impurity concentration. The LBTE method is used to scale the
scattering self-energies of NEGF, which only include local
interactions. This ensures an accurate reproduction of the measured
mobilities by NEGF. We then perform device simulations and demonstrate
that the considered transistors operate far from their performance
limit (from 50\% for MoS$_2$ to 60\% for WS$_2$). Higher quality
materials and substrate engineering will be needed to improve the
situation.
\end{abstract}

\section{Introduction}
Motivated by the first demonstration of a mono-layer MoS$_2$
transistor \cite{kis} significant efforts have been undertaken to
simulate the properties of devices based on 2-D transition metal
dichalcogenides (TMDs). Their carrier mobility has been extensively
modeled, starting with Ref.~\cite{kaas}: the phonon-limited mobility
of MoS$_2$ was computed with the linearized Boltzmann Transport
Equation (LBTE) after extracting the required effective masses and
electron-phonon coupling parameters from density-functional theory
(DFT). Going one step further, the influence of charged impurity
scattering was highlighted in Ref.~\cite{fischetti}, that of surface
optical phonons in Ref.~\cite{jena}, in both cases in the effective
mass approximation.

Accurately predicting the ``current vs. voltage'' characteristics of
monolayer TMD transistors represents another challenge. It can be
addressed with an \textit{ab initio} quantum transport (QT) approach,
e.g. the Non-equilibrium Green's Functions (NEGF), expressed in a
maximally localized Wannier function (MLWF) basis. As such
calculations are computationally very demanding, they were first 
restricted to ballistic transport \cite{chang}. Electron-phonon
scattering was later added, but in the form of parameterized
self-energies \cite{sengupta}. Attempts were also made to treat the
electrons, phonons, and their interactions at the
\textit{ab initio} level \cite{iedm16}. However, because of the
diagonal approximation typically applied to the scattering
self-energies, the obtained results may suffer from inaccuracies.

Here, we combine LBTE and NEGF, both from first-principles, to provide
the first QT simulations of MoS$_2$ and WS$_2$ single-layer transistors
with the same mobility values as in experiments. By considering
electron-phonon (E-Ph), charged impurity (CI), and surface optical
phonon (SOP) scattering, LBTE can reproduce the experimental mobility of
these 2-D materials over a large temperature range. These results were
then used to scale each diagonal scattering self-energy in NEGF so
that our QT calculations produce the same mobilities as LBTE. The
simulated ``I-V'' characteristics finally suggest that CI and SOP,
whose magnitude can be altered by improving the 2-D crystal quality or
by changing the substrate material, are responsible for a current
reduction by a factor comprised between 1.7 (WS$_2$) and 2 (MoS$_2$),
as compared to the case with E-Ph only (performance limit).

\section{Approach}\label{sec:app}
\begin{figure*}
\centering
\includegraphics[width=17.5cm]{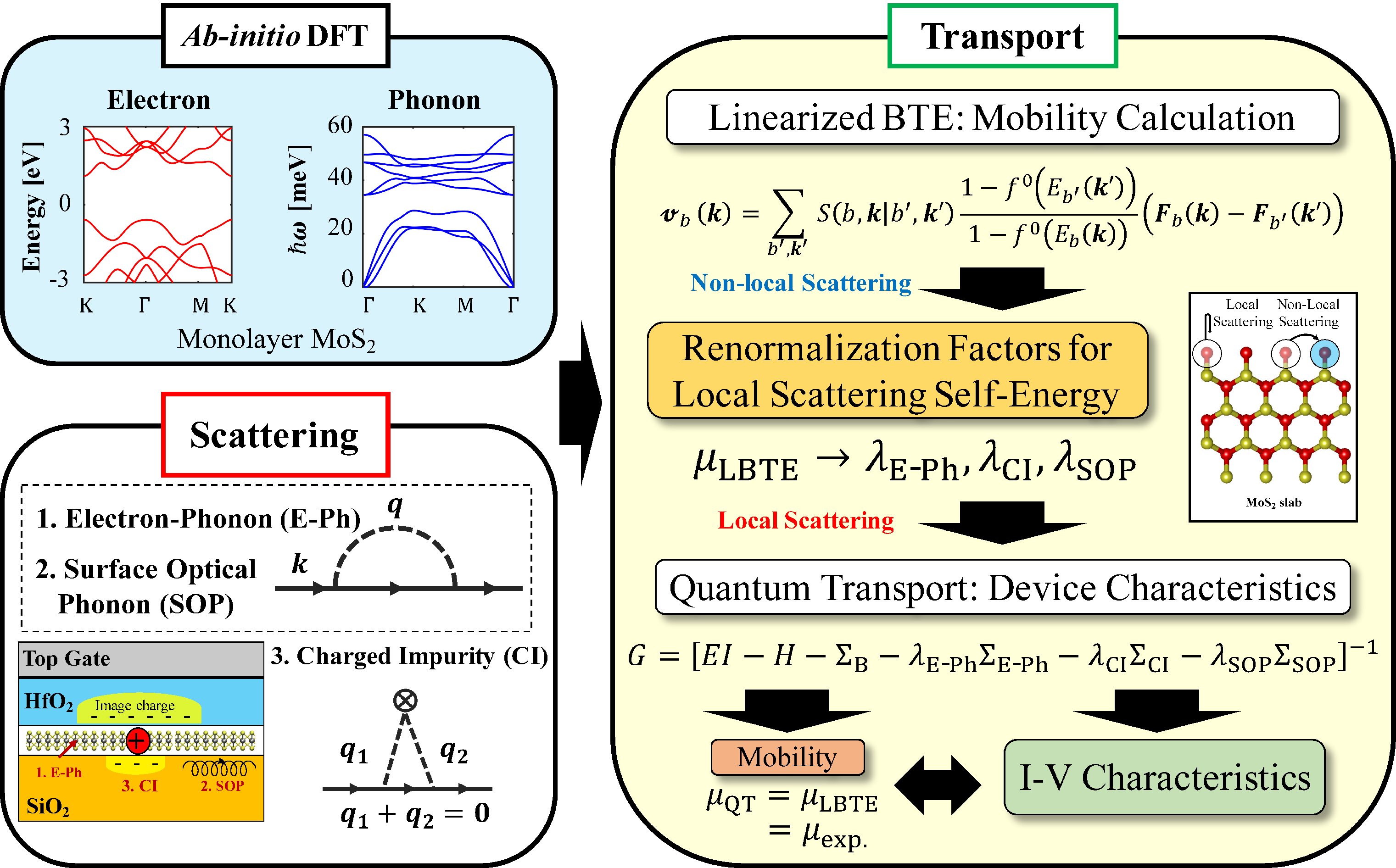}
\caption{Flowchart of the \textit{ab initio} simulation framework
  combining LBTE and quantum transport with NEGF. First, DFT is used
  to construct the Hamiltonian (electron) and dynamical (phonon)
  matrices of monolayer MoS$_2$ and WS$_2$ transistor
  structures. NEGF scattering self-energies ($\Sigma(E,k)$) and LBTE
  transition probability functions ($S(b,\mathbf{k}|b',\mathbf{k}')$)
  are then prepared for each interaction type (E-Ph, SOP, and CI)
  based on the corresponding Feynman diagrams. Screening is taken into
  account for SOP and CI. Mobility values are first computed with
  \textit{ab initio} LBTE, adjusting only the charged impurity
  concentration in order to reproduce experimental data over
  temperature. By comparing each mobility contribution ($\mu_{E-Ph}$,
  $\mu_{SOP}$, and $\mu_{CI}$) obtained with LBTE to its NEGF
  counterpart, renormalization factors $\lambda_{E-Ph}$,
  $\lambda_{SOP}$, and $\lambda_{CI}$ can be determined for each local
  scattering self-energy. After scaling the $\Sigma$'s with the
  $\lambda$'s, the NEGF quantum transport solver produces the same
  mobilities as LBTE and experiment and can therefore accurately
  predict device characteristics.}
\label{fig:1}
\end{figure*}

Our \textit{ab initio} LBTE+QT approach is summarized in
Fig.~\ref{fig:1}. As first step, the electronic structures of the
primitive unit cell corresponding to a 2-D monolayer were computed with 
VASP \cite{vasp} within the generalized gradient approximation (GGA)
of Perdew, Burke, and Ernzerhof (PBE) \cite{gga}. For transport
calculations, the resulting plane-wave DFT results were projected
onto a MLWF basis \cite{wannier} to produce a tight-binding-like
Hamiltonian \cite{szabo}. The dynamical matrices required for the
phonon frequencies were generated via density-functional perturbation
theory (DFPT) and the Phonopy code \cite{phonopy}. 

For each interaction type (E-Ph, CI, and SOP), a specific
scattering self-energy $\Sigma(E,k)$ (NEGF) and transition probability
function $S(b,\mathbf{k}|b',\mathbf{k}')$ (LBTE) was constructed from
the prepared DFT data and the corresponding Feynman diagram. In case
of E-Ph, the equations of Refs.~\cite{szabo} and \cite{rhyner} were
implemented for NEGF and LBTE, respectively. The electron-SOP coupling
was modeled by taking into account the vibrations of the dielectric
environments surrounding the 2-D monolayers \cite{jena}. The phonon
frequencies and coupling strengths were taken from measurements
\cite{sop}. Finally, the electron-CI interactions were cast into a
Coulomb potential with image charges \cite{fischetti} using the
impurity concentration $n_{imp}$ as fitting parameter. 2-D screening
caused by the surrounding dielectrics was applied to damp SOP and CI
scattering.

For each 2-D material, three mobilities $\mu_{E-Ph}$, $\mu_{SOP}$,
and $\mu_{CI}$ were calculated with the modified LBTE solver of 
Ref.~\cite{rhyner}, accounting for local and non-local energy-momentum 
transitions. Available temperature-dependent experimental data were
then reproduced by adjusting $n_{imp}$ only. Next, the same
mobilities were computed with NEGF using the ``dR/dL'' method
\cite{rim} and assuming diagonal scattering self-energies. By 
comparing the values obtained with LBTE and NEGF, renormalization
factors ($\lambda_{E-Ph}$, $\lambda_{SOP}$, and $\lambda_{CI}$) could
be determined to scale $\Sigma_{E-Ph}(E,k)$, $\Sigma_{SOP}(E,k)$, and
$\Sigma_{CI}(E,k)$, ensuring that NEGF returns the same mobility as
LBTE and experiments. Taking advantage of the ``calibrated'' NEGF
simulator \cite{szabo}, the I-V characteristics of 2-D devices could
be accurately predicted. 

\section{Results}\label{sec:res}
\begin{figure*}
\centering
\includegraphics[width=17.5cm]{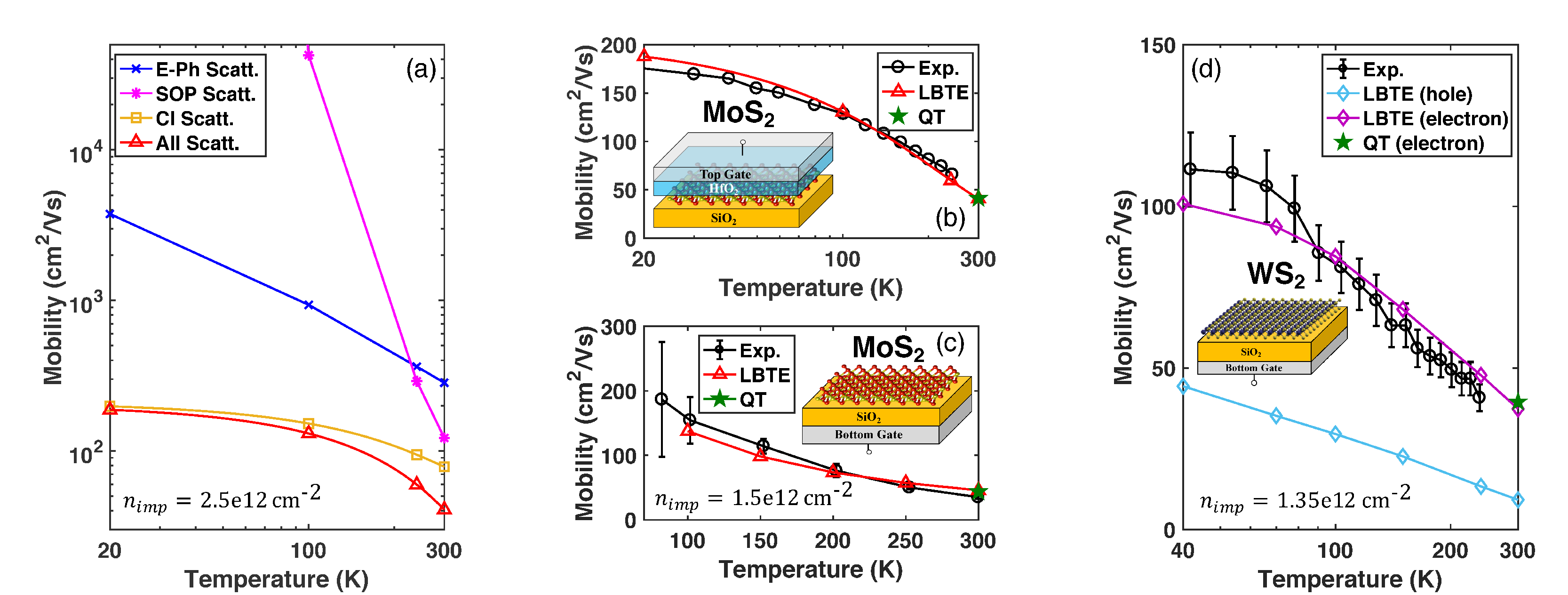}
\caption{(a) Temperature-dependent electron mobility contributions as
  calculated by LBTE. The E-Ph- ($\mu_{E-Ph}$, blue line
  with crosses), SOP- ($\mu_{SOP}$, pink line with asterisks), and
  CI-limited ($\mu_{CI}$, yellow line with squares) values are shown,
  together with the total mobility (red line with triangles) resulting
  from Matthiessen's rule. A monolayer MoS$_2$ with a 30 nm-thick top
  HfO$_2$ ($\epsilon_{ox}$=20) layer and a 270 nm-thick bottom SiO$_2$
  ($\epsilon_R$=3.9) substrate was used for that purpose. (b)
  Comparison of the temperature-dependent electron mobility calculated
  by LBTE (red line with triangles) to experimental data (black line
  with circles) \cite{mos2_1} for the same MoS$_2$ device as in
  (a) with $n_{imp}$=2.5e12 cm$^{-2}$. The green star refers to the
  NEGF mobility after scaling the scattering self-energies. (c) Same
  as (b), but for a MoS$_2$ monolayer deposited on a 270 nm-thick
  SiO$_2$ substrate with $n_{imp}$=1.5e12 cm$^{-2}$ \cite{mos2_2}. (d)
  Same as (b) and (c), but for a WS$_2$ monolayer on a 270 nm-thick
  SiO$_2$ substrate with $n_{imp}$=1.35 cm$^{-2}$ \cite{ws2} (black
  line with circles: experiment, lines with diamonds: LBTE).}
\label{fig:2}
\end{figure*}

The low-field mobility of MoS$_2$ and WS$_2$ monolayers as a function
of temperature is reported in Fig. \ref{fig:2}. Three cases were
investigated: MoS$_2$ deposited on a 270 nm-thick bottom SiO$_2$
substrate (i) with  and (ii) without a 30 nm-thick top HfO$_2$ layer
and (iii) WS$_2$ with the same substrate as (ii). For the MoS$_2$
configuration (i), the temperature-dependence of the E-Ph-, SOP-, and
CI-limited mobilities, as obtained with LBTE, is plotted in 
Fig.~\ref{fig:2}(a), together with the total mobility calculated by
Matthiessen's rule. The impurity concentration was set to
$n_{imp}$=2.5e12 cm$^{-2}$, which leads to an excellent agreement with
the experimental data of Ref.~\cite{mos2_1} (Fig.~\ref{fig:2}(b)). It
can be seen that CI scattering is the limiting factor at all
temperatures, but SOP kicks in at high temperatures and induces a
significant mobility drop. The LBTE mobilities of the second MoS$_2$
sample (Fig.~\ref{fig:2}(c), $n_{imp}$=1.5e12 cm$^{-2}$) \cite{mos2_2}
and of WS$_2$ (Fig.~\ref{fig:2}(d), $n_{imp}$=1.35e12 cm$^{-2}$)
\cite{ws2} reveal an equally good agreement with experiments. In
Fig.~\ref{fig:2} we also provide the room-temperature QT mobilities
with renormalized scattering self-energies to validate our scheme. 

\begin{figure*}
\centering
\includegraphics[width=17.5cm]{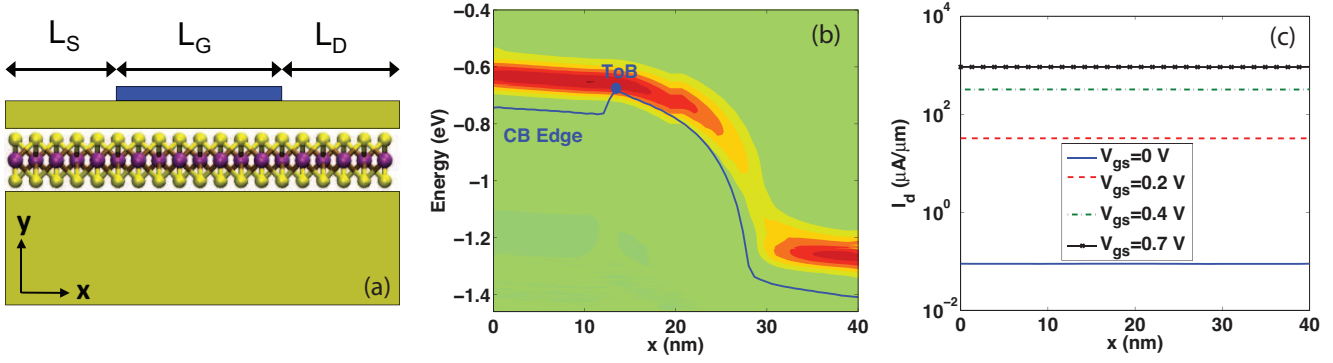}
\caption{(a) Schematic view of the single-gate (SG) monolayer MoS$_2$ and 
  WS$_2$ field-effect transistors (FETs) investigated here. The source
  and drain extensions measure $L_s$=$L_d$=15 nm each and are doped
  with a donor concentration $N_D$=5e13 cm$^{-2}$. Ohmic contacts are
  assumed. The gate length $L_g$ is set to 15 or 50 nm. The 2-D
  materials lie on a 20 nm thick SiO$_2$ ``substrate'' with
  $\epsilon_R$=3.9. The gate contact is separated from the channel by 
  a $t_{ox}$=3 nm HfO$_2$ layer with $\epsilon_{ox}$=20. All device
  simulations were performed at $V_{ds}$=0.7 V and at room
  temperature. (b) Spectral distribution of the current through the
  WS$_2$ FET with $L_g$=15 nm, $V_{gs}$=$V_{ds}$=0.7 nm, and in the 
  presence of all scattering mechanisms considered in this work. Red
  indicates high current concentrations, green none. The blue line
  refers to the conduction band edge of WS$_2$. (c) Spatial current
  distribution of the same WS$_2$ device as in (b) as a function of
  $V_{gs}$ and the position along the $x$ transport direction. All
  scattering sources discussed in this paper are included.} 
\label{fig:3}
\end{figure*}

As next step, the ``current vs. voltage'' characteristics of MoS$_2$
and WS$_2$ monolayer field-effect transistors (FETs) are analyzed. The
selected benchmark structure is shown in Fig.~\ref{fig:3}(a): a
single-gate (SG) FET with a gate length $L_g$=15 or 50 nm,
source/drain-extensions of $L_s$=$L_d$=15 nm each, and a supply
voltage $V_{DD}$=0.7 V. The 2-D materials were deposited on a SiO$_2$
``substrate" with a top HfO$_2$ layer of thickness $t_{ox}$=3 nm and a
relative permittivity $\epsilon_{ox}$=20. Ohmic contacts were assumed
with a donor concentration $N_D$=5e13 cm$^{-2}$. Hence, the effect of
contact resistances was neglected. To verify that the QT scattering
models properly work, the energy- and position-resolved current of
the WS$_2$ FET is depicted in Fig.~\ref{fig:3}(b), including E-Ph,
SOP, and CI interactions, while current conservation is demonstrated
in Fig.~\ref{fig:3}(c) for different gate biases. As expected,
electrons lose a substantial amount of their energy on the drain side
of the transistor, where the electric field reaches its maximum.

\begin{figure*}
\centering
\includegraphics[width=17.5cm]{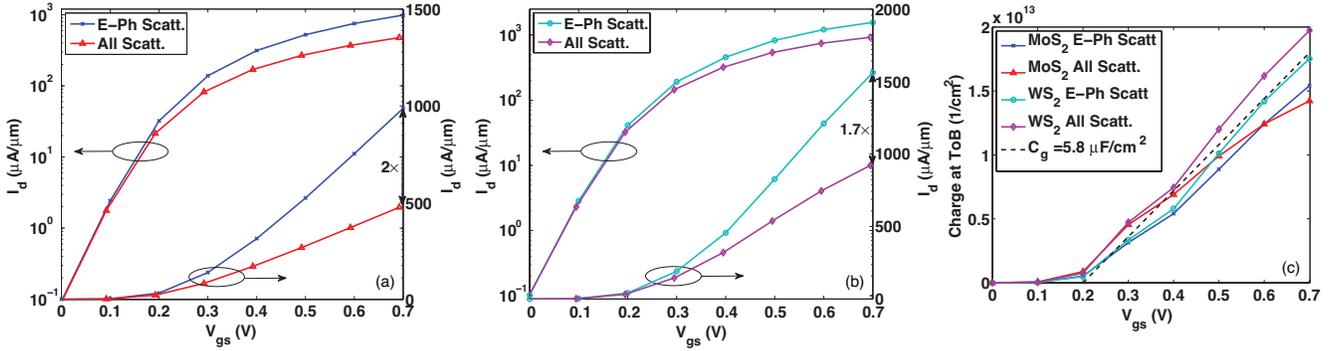}
\caption{(a) $I_d$-$V_{gs}$ transfer characteristics of the monolayer
  MoS$_2$ single-gate FET plotted on a linear (right $y$-axis) and on a
  logarithmic (left $y$-axis) scale. The current with electron-phonon
  scattering only (blue lines with crosses) and with all scattering
  sources (red lines with triangles) are reported at $V_{ds}$=0.7 V,
  $I_{OFF}$=0.1 $\mu$A/$\mu$m, and with a charged impurity
  concentration $n_{imp}$=2.5e12 cm$^{-2}$. (b) Same as (a), but for
  the WS$_2$ FET and $n_{imp}$=1.35e12 cm$^{-2}$. (c) Electron
  concentration extracted at the top-of-the-barrier (ToB, see
  Fig.~\ref{fig:3}) as a function of $V_{gs}$ for the four cases
  presented in (b) and (c). An average gate capacitance $C_g$=5.8
  $\mu$F/cm$^2$ can be extracted from these results (dashed black
  line), closed to the oxide capacitance
  $C_{ox}$=$t_{ox}$/$\epsilon_{ox}$=5.9 $\mu$F/cm$^2$.}
\label{fig:4}
\end{figure*}

The $I_{d}$-$V_{gs}$ transfer characteristics of both FET types are
shown in Figs.~\ref{fig:4}(a-b). Two curves per device were
computed. First, as a reference, the current with electron-phonon
scattering only is given. This mechanism cannot be eliminated, 
contrary to CI that can be reduced by improving the crystal quality or
SOP that could possibly be minimized by carefully choosing the
substrate material. The second curve includes all scattering sources
and uses the impurity concentrations and scaling factors from
Fig.~\ref{fig:2}. A strong ON-current decrease caused by SOP and CI is
found. It is larger in MoS$_2$ than in WS$_2$ (2 vs. 1.7$\times$) due 
to the presence of more charged impurities ($n_{imp}$=2.5 vs. 1.35
cm$^{-2}$). Consequently, both FETs operate far from their optimum,
defined as the current with E-Ph scattering only. It can be noticed in
Fig.~\ref{fig:4}(c) that the gate capacitance $C_g$=5.8
$\mu$F/cm$^{2}$ extracted from the derivative of the electron
concentration at the top-of-the-barrier with respect to the gate
voltage does not suffer from so-called density-of-states bottlenecks
as it approaches the value of $C_{ox}$=$\epsilon_{ox}$/$t_{ox}$=5.9
$\mu$F/cm$^{2}$, the oxide capacitance.  

\begin{figure*}
\begin{minipage}{0.66\linewidth}
\centering
\includegraphics[width=\linewidth]{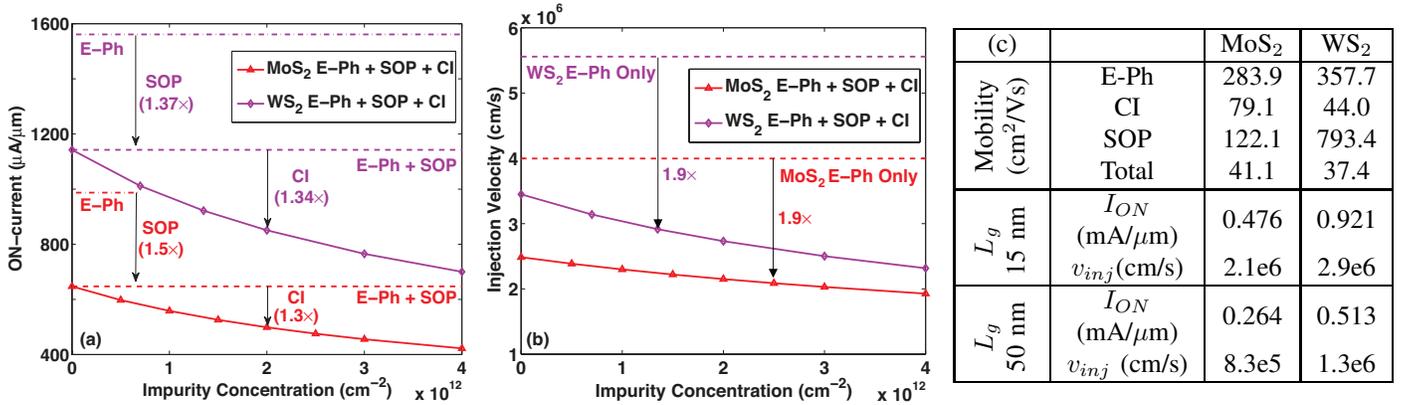}
\end{minipage}
\hspace{0.1cm}
\begin{minipage}[b]{0.33\linewidth}
\begin{tabular}{|c|c|c|c|}
\hline
(c)&&MoS$_2$&WS$_2$\\
\hline
\multirow{4}{*}{\STAB{\rotatebox[origin=c]{90}{\makecell{Mobility\\(cm$^2$/Vs)}}}}&E-Ph&283.9&357.7\\
&CI&79.1&44.0\\
&SOP&122.1&793.4\\
&Total&41.1&37.4\\
\hline
\multirow{2}{*}{\STAB{\rotatebox[origin=c]{90}{\makecell{$L_g$ \\15 nm}}}}&\makecell{$I_{ON}$\\(mA/$\mu$m)}&0.476&0.921\\
&$v_{inj}$(cm/s)&2.1e6&2.9e6\\
\hline
\multirow{2}{*}{\STAB{\rotatebox[origin=c]{90}{\makecell{$L_g$ \\50 nm}}}}&\makecell{$I_{ON}$\\(mA/$\mu$m)}&0.264&0.513\\
&$v_{inj}$ (cm/s)&8.3e5&1.3e6\\
\hline
\end{tabular}
\end{minipage}
\caption{(a) ON-current at $I_{OFF}$=0.1 $\mu$A/$\mu$m as a function
  of the charged impurity concentration for the MoS$_2$ (red line with
  triangle) and WS$_2$ (purple line with diamonds) single-gate
  FET. E-Ph and SOP scattering are included as well. The dashed lines
  refer to the E-Ph+SOP (no CI) case, the dashed-dotted lines to E-Ph
  only. The current reductions caused by SOP and CI (at $n_{imp}$=2e12
  cm$^{-2}$) are indicated by the arrows. (b) Electron injection
  velocity at the ToB location corresponding to the ON-currents in
  (a). (c) Table summarizing the calculated room temperature mobility
  values (E-Ph, CI, SOP, and total) at an electron concentration 1e13
  cm$^{-2}$ (4.8e12 cm$^{-2}$) for MoS$_2$ (WS$_2$) and $I$-$V$
  characteristics of the corresponding FETs with $n_{imp}$=2.5e12
  cm$^{-2}$ (1.35e12 cm$^{-2}$). The ON-current and injection velocity
  at the ToB are provided for devices with two different gate lengths
  ($L_g$=15 and 50 nm).} 
\label{fig:5}
\end{figure*}

The influence of SOP and CI is examined in Fig.~\ref{fig:5}(a) where
the ON-current values of the MoS$_2$ and WS$_2$ FETs are plotted as a
function of the impurity concentration $n_{imp}$. The cases with E-Ph
scattering only, E-Ph+SOP, and E-Ph+SOP+CI are supplied, thus
isolating the impact of each scattering mechanism from the others. It
appears that SOP contributes slightly more to the current decrease
than CI (factor 1.37-1.5$\times$ vs. 1.3-1.34$\times$), for both 2-D
materials, contrary to what was observed in mobility calculations
(CI-limited). This seems to indicate that the emission/absorption of
SOP is enhanced under high-field conditions. The corresponding
injection velocities at the top-of-the-barrier follow in
Fig.~\ref{fig:5}(b). Thanks to their lower effective mass
($m_e^*$=0.30 vs. 0.46 $m_0$), electrons in WS$_2$ are faster than in
MoS$_2$, but their velocity could still be improved by a factor
$\sim$2, as compared to the present case, if both SOP and CI were
removed. The extracted velocities however degrade when the gate length
increases to 50 nm, as can be seen in Table \ref{fig:5}(c). The 
latter summarizes the key mobility, ON-current, and injection velocity
results obtained here.  

\section{Conclusion}\label{sec:conc}
We proposed an \textit{ab initio} LBTE+QT framework to compute the
transport properties of 2-D materials. It relies on the
renormalization of the NEGF self-energies with scaling factors
determined from LBTE. The mobility of MoS$_2$ and WS$_2$ monolayers
was accurately modeled with this method, which also helped predict the
``I-V'' characteristics of monolayer MoS$_2$ and WS$_2$ FETs with
short gate lengths. Further device progresses will depend on improved
materials and careful design choices. The developed environment is now
ready to explore novel 2-D compounds beyond TMDs.  

\section*{Acknowledgment}
This work was supported the MARVEL National  Centre  of  Competence
in  Research  of  the  Swiss  National Science Foundation (SNSF), by
SNSF under Grant No.~175479 (ABIME), and by a grant from the Swiss
National Supercomputing Centre (CSCS) under Project s876.

\bibliographystyle{IEEEtran}

\newpage

\newpage

\end{document}